\documentclass[aps,preprint,prl,byrevtex]{revtex4-1}
\usepackage{amssymb}
\usepackage{amsfonts}
\usepackage{amsmath}
\usepackage{graphicx}
\begin{document}
\preprint{HEP/123-qed}
\title[Short title for running header]{Persistence of superconducting condensate far above the critical temperature in YBa$_2$(Cu,Zn)$_3$O$_y$ revealed by \textit{c}-axis optical conductivity measurements for several Zn concentrations and carrier doping levels}
\author{Ece Uykur}
\affiliation{Department of Physics, Graduate School of Science, Osaka University, 560-0043,
Osaka, JAPAN}
\author{Kiyohisa Tanaka}
\affiliation{Department of Physics, Graduate School of Science, Osaka University, 560-0043,
Osaka, JAPAN}
\author{Takahiko Masui}
\affiliation{Department of Physics, Graduate School of Science, Osaka University, 560-0043,
Osaka, JAPAN}
\author{Shigeki Miyasaka}
\affiliation{Department of Physics, Graduate School of Science, Osaka University, 560-0043,
Osaka, JAPAN}
\author{Setsuko Tajima}
\affiliation{Department of Physics, Graduate School of Science, Osaka University, 560-0043,
Osaka, JAPAN}
\pacs{PACS number}

\begin{abstract}
The superconductivity precursor phenomena in high temperature cuprate
superconductors was studied by direct measurements of superconducting
condensate with the use of the \textit{c}-axis optical conductivity of
YBa$_{2}$(Cu$_{1-x}$Zn$_{x}$)$_{3}$O$_{y}$ for several doping levels
($p$) as well as for several Zn-concentrations. Both of the real and imaginary
part of the optical conductivity clearly showed that the superconducting
carriers persist up to the high temperatures \textit{T}$_{p}$ that is higher
than the critical temperature \textit{T}$_{c}$ but lower
than the pseudogap temperature \textit{T}$^{\ast}$. \textit{T}$_{p}$ increases
with reducing doping level like \textit{T}$^{\ast}$, but decreases with
Zn-substitution unlike \textit{T}$^{\ast}$. 

\end{abstract}
\volumeyear{year}
\volumenumber{number}
\issuenumber{number}
\eid{identifier}
\maketitle

Precursor superconductivity is one of the subjects that currently attract a
lot of attention in the research of high temperature superconducting cuprates
\cite{Wang2006, Li2010, Kondo2011, Bilbro2011, Grbic2011, Dubroka2011}.
Despite the number of reports presented so far, this problem is still
controversial. One common observation is that the superconducting fluctuation
regime can be described in a temperature range between \textit{T}$_{c}$ (superconducting critical temperature) and
\textit{T}$^{\ast}$ (pseudogap temperature). However, the carrier-doping dependence and the temperature range
show significant differences among the different probes. Some experiments like
Nernst effect \cite{Wang2006}, diamagnetism measurements \cite{Li2010}
reported the superconducting fluctuation regime at the temperatures as high as
3-4 \textit{T}$_{c}$ with a doping dependence different from that of
\textit{T}$_{c}$. On the other hand, the microwave experiments
\cite{Grbic2011} showed that the temperature range of the superconducting
fluctuation is only 10-20 K above \textit{T}$_{c}$ with a doping dependence
similar to the \textit{T}$_{c}$ dome.

Here we report the observation of superconducting condensate above
\textit{T}$_{c}$ through the \textit{c}-axis optical spectra of Zn-doped YBa$_2$Cu$_3$O$_y$ (YBCO) single
crystals for several doping levels ($p$) in a wide energy range ($2.5$ meV --
$40$ eV ). Doping level $p$ in YBCO varies with the oxygen concentration \cite{Liang2006} and the maximum \textit{T}$_{c}$ is achieved at $p$ = 0.16 (optimum doping). In the \textit{c}-axis optical conductivity, we can clearly
distinguish the pseudogap and the superconducting gap, while in many probes it
is difficult to distinguish these two gaps. YBCO has high conductivity along
the \textit{c}-axis, compared to many other cuprates, which allows us to
detect small changes of conductivity. Zn-substitution, on the other hand, is
useful to resolve the superconductivity related responses. Temperature
dependent reflectivity measurements are performed on YBa$_{2}$(Cu$_{1-x}
$Zn$_{x}$)$_{3}$O$_{y}$ single crystals, where the details regarding the
samples are described in the supplementary information. 

The \textit{c}-axis optical conductivity is suppressed at low energies both by
the pseudogap and the superconducting gap, while the direction of the spectral
weight (SW) transfer is opposite in these two gaps. The lost SW is transferred
to the high energy region for the pseudogap, while it is redistributed at
$\delta$-function at zero frequency for the superconducting gap. We determined
the pseudogap opening temperature \textit{T}$^{\ast}$ by tracing the
suppression of the low energy optical conductivity (at 20 cm$^{-1}$), as shown
in Fig. 1. When the temperature decreases from room temperature, the low
energy optical conductivity gradually increases due to the metallic response
of the system, and below \textit{T}$^{\ast}$, it begins to decrease. The
metallic behavior is weakened and the turning point - \textit{T}$^{\ast}$ -
increases with underdoping. The pseudogap issue has been discussed in many
experiments \cite{Timusk1999}. Our \textit{T}$^{\ast}$ values for the Zn-free
samples are in good agreement with those by the other methods such as the
measurements of the $^{89}$Y nuclear magnetic resonance (NMR) Knight shift
\cite{Alloul2012} and resistivity \cite{Daou2010}. Moreover, as it has been
pointed out by many research groups \cite{Alloul1991, Mizuhashi1995,
Yamamoto2002, Uykur2013}, the pseudogap temperature does not change
significantly with Zn-substitution (see Fig. 1b and 1c). 

\begin{figure}[h]
\centering
\includegraphics[
trim=0.000000in 0.000000in 0.000764in 0.000000in,
height=1.3979in,
width=3.4634in
]
{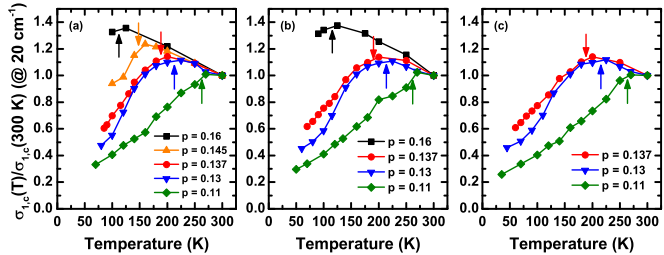}
\caption{$T$-dependence of low energy optical conductivity for various $p$
with Zn-content $x$ = 0 (a), $x$ = 0.007 (b), and $x$ = 0.012 (c). The arrows
indicate the pseudogap temperature, $T$*. All the values are normalized to the
room temperature value.}
\label{F1}
\end{figure}

\begin{figure}[ptbh]
\centering
\includegraphics[
height=6.3323in,
width=5.1859in
]
{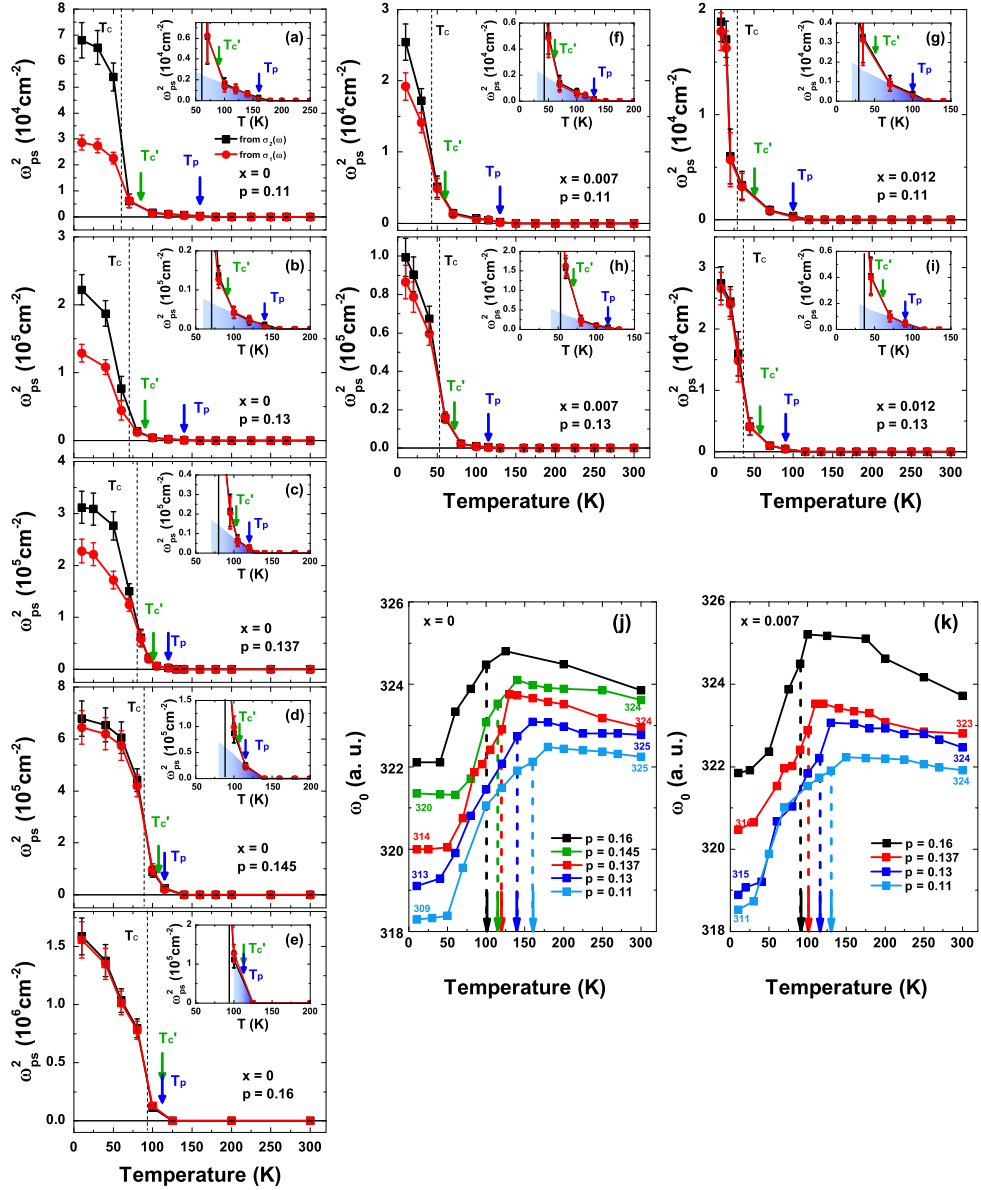}
\caption{Determination of the precursor superconducting state. (a) - (g) are
the temperature dependences of $\omega_{ps}^{2}$ obtained from the missing
area in $\sigma_{1}(\omega)$ (red cicles) and from $\sigma_{2}(\omega)$ (black
squares). Insets are the expanded figures near \textit{T}$_{p}$. Figures (a)
to (e) demonstrate the doping dependent behavior for \textit{x} = 0, while in
Figs. (a, f, g) and Figs. (b, h, i) we can see the \textit{x}-dependence at
\textit{p} = 0. 11 and 0.13, respectively. (j) and (k) The resonance
frequencies of the oxygen bending mode for \textit{x} = 0 and 0.007,
respectively. Error bars in these figures are discussed in the supplementary
information.}
\end{figure}

Optical probe is also very sensitive to the superconducting carrier response. The imaginary part of the optical conductivity ($\sigma_{2}(\omega)$) is
directly related to the superfluid density ($4\pi\omega\sigma_{2}(\omega)$ =
$\omega_{ps}^{2}$ = $4\pi n_{s}/m^{\ast}$). Another way to calculate the
superfluid density from optical spectra is to estimate the missing area in the
real part of the optical conductivity ($\sigma_{1}(\omega)$). Since we found
that the spectral weight of $\sigma_{1}(\omega)$ is conserved below 5000
cm$^{-1}$ in our previous study \cite{Uykur2013}, the missing area was
estimated from the deviation from this conserved value for each sample. In
Fig. 2, we plot the temperature dependence of $\omega_{ps}^{2}$ determined
from the missing area in $\sigma_{1}(\omega)$ (red circles) and from
$\sigma_{2}(\omega)$ (black squares) for various doping levels and
Zn-contents. Details of the data analysis can be found in the supplementary
information. The results by both methods are qualitatively the same: zero
values at high temperatures, followed by a rapid increase at \textit{T}$_{c}$
due to the superconducting transition. A closer look at each figure (the insets) revealed a finite value of the superfluid densities ($\omega
_{ps}^{2}$) at a certain temperature higher than
\textit{T}$_{c}$. We name this temperature the precursor superconductivity temperature \textit{T}$_{p}$. For the underdoped Zn-free sample (\textit{p} = 0.11), this
value is as high as 160 K. With decreasing temperature, $\omega_{ps}^{2}$
gradually increases and the slope of increase suddenly becomes steeper at the temperature near \textit{T}$_{c}$. Hereafter, we refer to this temperature as  \textit{T}$_{c}^{\prime}$. From all the figures, we can find that the superconducting carriers persist up to much higher temperatures than
\textit{T}$_{c}$, although its fraction is very small (less than a few \% of
the total $\omega_{ps}^{2}$ at \textit{T} = 0).

It is interesting that the doping dependences of \textit{T}$_{c}^{\prime}$ and
\textit{T}$_{p}$ are different (Figs. 2(a-e)). \textit{T}$_{c}^{\prime}$ is always
10-20 K above \textit{T}$_{c}$ and thus follow the \textit{T}$_{c}$ change,
whereas \textit{T}$_{p}$ increases with decreasing doping levels (\textit{p})
and reaches much higher temperatures than \textit{T}$_{c}^{\prime}$. \textit{T}$_{c}^{\prime}$
and \textit{T}$_{p}$ are almost merged at the optimum doping \textit{p} = 0.16
(Fig. 2(e)). Although these two temperature scales (\textit{T}$_{p}$ and
\textit{T}$_{c}^{\prime}$) show different doping dependences, the Zn-dependences are
similar (Figs. 2(a), 2(f), 2(g) for \textit{p} = 0.11 and Figs. 2(b), 2(h),
2(i) for \textit{p} = 0.13). Namely, as Zn content increases, both
\textit{T}$_{p}$ and \textit{T}$_{c}^{\prime}$ decrease just like \textit{T}$_{c}$
\cite{Maeno1987}.

The transverse Josephson plasma (TJP) resonance mode \cite{vanderMarel1996}
might be used as a measure of superconductivity, as well. A broad conductivity
peak starts to appear weakly above \textit{T}$_{c}$ in the far-infrared
region, and significantly grows below \textit{T}$_{c}$. Meanwhile, it strongly
couples with the oxygen bending mode phonon at $\sim$ 320 cm$^{-1}$, which
results in the change in the SW and the resonance frequency of this phonon
mode. Therefore, by tracing the phonon frequency ($\omega_{0}$) of the oxygen
bending mode, we can identify the temperature at which the TJP mode starts to
evolve. Dubroka \textit{et} \textit{al}. attributed the appearance of the TJP
resonance mode above \textit{T}$_{c}$ to the precursor superconductivity
effect \cite{Dubroka2011}. We also plot $\omega_{0}$ to estimate the onset
temperature of precursor superconductivity, assuming that the TJP mode is an
indication of superconductivity. From the temperature dependence of
$\omega_{0}$ (Figs. 2(j) and 2(k)), we can estimate the values of
\textit{T}$_{p}$ at which $\omega_{0}$ starts to decrease. In both the pure
(\textit{x} = 0) and the Zn-doped (\textit{x} = 0.007) samples, for all the
doping levels, \textit{T}$_{p}$ values obtained from the TJP resonance mode
coincide with those estimated from $\sigma_{2}(\omega)$ and $\sigma_{1}
(\omega)$. The fact that the three independent methods give the same
\textit{T}$_{p}$ values confirms that our observation of superconductivity up
to \textit{T}$_{p}$ is real.

We plot all the temperatures \textit{T}$^{\ast}$, \textit{T}$_{p}$,
\textit{T}$_{c}^{\prime}$ , and \textit{T}$_{c}$ for various doping levels and
Zn-contents in Fig. 3(a), where for each Zn-content (\textit{x})
\textit{T}$_{c}$ is normalized by the value at \textit{p} = 0.16 (namely, the
normalized \textit{T}$_{c}$ equals to 1 at \textit{p} = 0.16) and
\textit{T}$_{c}^{\prime}$, \textit{T}$_{p}$, and \textit{T}$^{\ast}$ are multiplied by
the same normalization factor for each \textit{x}. It turns out that
\textit{T}$_{c}^{\prime}$ and \textit{T}$_{p}$ for different \textit{x} samples form a
single curve, which indicates that both of these temperatures change with
\textit{x}, scaling with \textit{T}$_{c}$, unlike the pseudogap temperatures
(\textit{T}$^{\ast}$ does not change with \textit{x}). The Zn-dependence of
\textit{T}$_{p}$ and \textit{T}$_{c}^{\prime}$ gives further evidence that we really
observe the superconducting signature above \textit{T}$_{c}$. It is natural to
consider that \textit{T}$_{c}^{\prime}$ corresponds to a conventional superconducting
fluctuation temperature described by the Ginzburg-Landau formalism
\cite{Larkin2009}, but \textit{T}$_{p}$ indicates the unusual phenomenon due
to superconductivity precursor. These systematic Zn- and doping-dependent
behaviors prove that our observation of precursor superconductivity is neither
due to the sample inhomogeneity nor due to the measurement errors.

\begin{figure}[ptbh]
\centering
\includegraphics[
height=2.089in,
width=3.4634in
]
{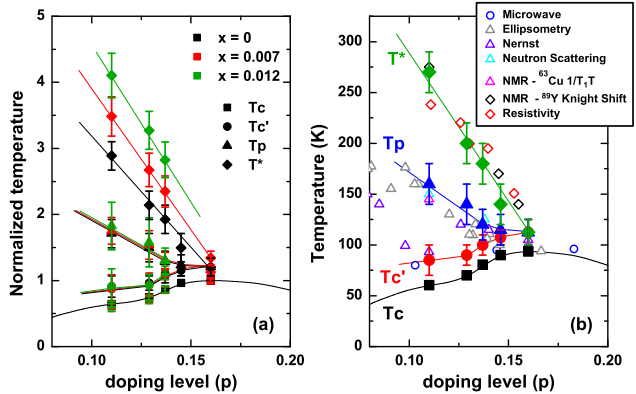}
\caption{Electronic phase diagram of YBa$_{2}$(Cu$_{1-x}$Zn$_{x}$)$_{3}$
O$_{y}$. (a) The normalized variable for \textit{T}$_{c}^{\prime}$ (circles),
\textit{T}$_{p}$ (triangles), and \textit{T}$^{\ast}$ (diamonds) for
\textit{x} = 0 (black symbols), \textit{x} = 0.007 (red symbols), and
\textit{x} = 0.012 (green symbols). (b) Phase diagram of the YBa$_{2}$Cu$_{3}
$O$_{y}$ where the temperatures obtained by our measurements (solid symbols)
are plotted together with the data by several other probes (Microwave
\cite{Grbic2011}, ellipsometry \cite{Dubroka2011}, Nernst effect
\cite{Ong2003}, neutron scattering \cite{Dai1999}, NMR $^{63}$Cu 1/T$_{1}$T
\cite{Takigawa1991, Zheng1996}, NMR $^{89}$Y Knight shift \cite{Alloul2012},
and resistivity \cite{Daou2010}.}
\label{F3}
\end{figure}

The other new finding in the present study is that the difference of
$\omega_{ps}^{2}$ estimated from $\sigma_{2}(\omega)$ and the missing area in
$\sigma_{1}(\omega)$ diminishes with Zn-substitution. As we see in Fig. 2, for
the Zn-free samples, although the temperature dependences of $\omega_{ps}^{2}$
are similar, the two estimation methods give significantly different values of
$\omega_{ps}^{2}$ in the underdoped regime. This discrepancy has been
previously pointed out \cite{Basov2001}, and attributed to the kinetic energy
reduction \cite{Hirsch1992}. In this scenario, not only the low energy SW but
also the high energy SW in the visible region are condensed into a delta
function at $\omega$ = 0.

A similar reduction of the discrepancy in $\omega_{ps}^{2}$ has been reported
in the optical measurements under magnetic fields for YBCO \cite{LaForge2009}.
The common effect of magnetic field and Zn-substitution is that the TJP
resonance mode is suppressed in both cases. The TJP mode can be seen as a
broad peak in $\sigma_{1}(\omega)$ below \textit{T}$_{c}$, which reduces the
calculated missing area \cite{vanderMarel1996, Munzar2003}. On the other hand,
the low energy $\sigma_{2}(\omega)$ is not affected by this mode. Therefore,
we conclude that the observed difference in $\omega_{ps}^{2}$ in the two
methods is caused by the inappropriate estimation of the missing area, i.e.
ignoring the contribution of the TJP mode to $\omega_{ps}^{2}$. In other
words, when the TJP mode is suppressed, the difference in $\omega_{ps}^{2}$
should vanish. Note that the TJP mode above \textit{T}$_{c}$ is very weak
\cite{Bernhard2000} and thus it does not couse an appreciable difference in
$\omega_{ps}^{2}$ in Fig. 2.

In Fig. 3(b) we compare our results for the Zn-free sample with the published
data of YBCO determined by the other probes. Solid symbols represent our data.
Our \textit{T}$_{c}^{\prime}$ values are in good agreement with the recent results of
microwave measurements on YBCO \cite{Grbic2011}. Moreover, the temperature
scale of \textit{T}$_{c}^{\prime}$ relative to \textit{T}$_{c}$ is also consistent
with the results of THz \cite{Bilbro2011, Nakamura2012, Corson1999} and microwave
measurements \cite{Ohashi2009} for the other cuprate systems, La$_{2-x}$Sr$_{x}$CuO$_{4}$ and Bi$_2$Sr$_2$CaCu$_2$O$_{8+\delta}$. On the other hand, neither THz nor microwave measurements detected
the temperature scale \textit{T}$_{p}$. This might be due to the ambiguity in
determining the normal carrier component which we need to subtract in the
analysis to calculate $\omega_{ps}^{2}$ from $\sigma_{2}(\omega)$
\cite{Dordevic2002}.

Our \textit{T}$_{p}$ values are in good agreement with the temperatures
observed by ellipsometry \cite{Dubroka2011} and partly by Nernst effect
\cite{Ong2003, explanation}. In Ref. 6, \textit{T}$_{p}$ was estimated only
from the phonon softening related to the TJP resonance, which is not direct
evidence for superconducting condensate. Moreover, we cannot adopt this method
to follow the \textit{T}$_{p}$ change with Zn-substitution, since the TJP mode
is gradually suppressed with Zn-substitution. It is worth noting that
\textit{T}$_{p}$ well coincides with the spin gap temperature reported by
neutron scattering \cite{Dai1999} and the relaxation rate T$_{1}^{-1}$ of NMR
\cite{Takigawa1991, Zheng1996}.

Our results indicates that a precursor of superconductivity does exist at
temperatures much higher than \textit{T}$_{c}$ but lower than \textit{T}
$^{\ast}$. This precursor phenomenon is clearly distinguished from the
pseudogap not only because of the difference in temperature scale but also
because of the fact that the electrons removed from the Fermi surface owing to
the pseudogap never contribute to superconductivity \cite{Uykur2013, Yu2008}.
Moreover, the observation of superconducting condensate implies that the
Cooper pairs are formed with phase coherence at \textit{T}$_{p}$.

These experimental facts put a strong constraint on the theory for
high-\textit{T}$_{c}$ superconductivity. For example, preformed pairs
predicted by the mean field theory of \textit{t-J} model \cite{Lee2006} do not
have phase coherence and thus they cannot explain our observation.
Microscopically phase separated state in a doped Mott insulator
\cite{Kresin2006, Emery1995, Emery1997} is a more plausible candidate.
Recently the charge density wave (CDW) order was observed in the underdoped
YBCO at the temperature close to our \textit{T}$_{p}$ \cite{Ghiringhelli2012,
Chang2012}. The simultaneous observation of a precursor of superconductivity
and the CDW order suggests microscopic phase separation. Moreover, we may
expect some interplay between these two orders although they originally
compete with each other. To discuss the origin of this unusual precursor, the
increase of \textit{T}$_{p}$ with decreasing the doping level is a smoking gun
pointing to the importance of Mottness in the high-\textit{T}$_{c}$
superconductivity mechanism.

\bigskip

\begin{acknowledgments}
We thank T. Tohyama and D. Van der Marel for the useful discussion. This work
was supported by the Grant-in-Aid for Scientific Research from the Ministry of
Education, Culture, Sports, Science and Technology of Japan (KIBAN(A)
No.19204038 and KIBAN(B) No. 24340083).
\end{acknowledgments}

\section{SUPPLEMENTARY}

\subsection{I. Samples}

The high quality YBa$_{2}$(Cu$_{1-x}$Zn$_{x}$)$_{3}$O$_{7-\delta}$ single
crystals with several Zn-contents ($x$) were grown by the pulling technique
[1]. The doping level ($p$) was varied by adjusting the oxygen concentrations
and determined from the experimental $p$\textit{ - T}$_{c}$ curve [2]. The
Zn-free and the Zn-substituted samples were simultaneously annealed at a
specific temperature for a specific doping level under oxygen flow, then
rapidly quenched into liquid nitrogen. The superconducting transition
temperatures (\textit{T}$_{c}$) were determined from the dc susceptibility
measurements. The susceptibility curves for all the samples are given in Fig.
S1 (a)-(c) for \textit{x} = 0, 0.007, and 0.012, respectively. All the samples
show rather sharp, single phase transitions. The annealing conditions, doping
levels and \textit{T}$_{c}$ values with transition width $\Delta$
\textit{T}$_{c}$ are summarized in Table I.

\begin{figure}[ptbh]
\centering
\includegraphics[
height=2.3112in,
width=5.7562in
]
{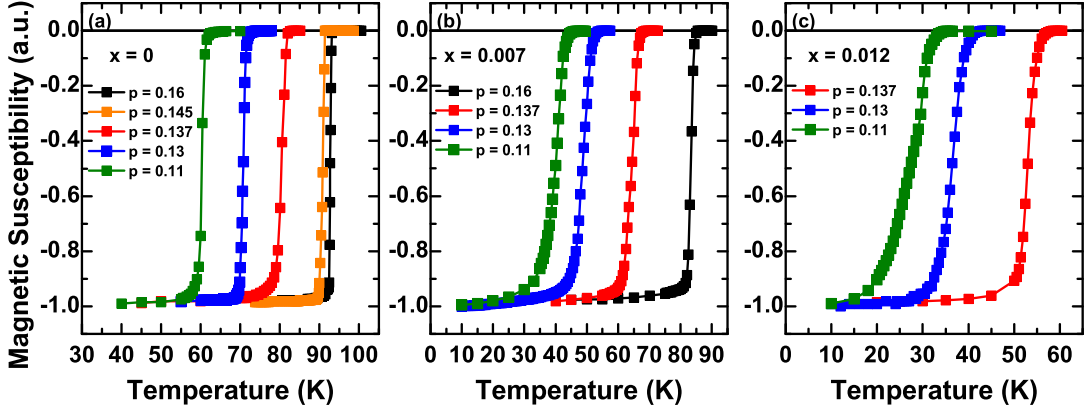}
\caption{Magnetic susceptibility vs temperature curves of the (a) $x=0$, (b) $x=0.007$, and (c) $x=0.012$ for several doping
levels ($p$). Magnetizations are normalized at the lowest temperature
($T=10$ K).}
\label{FS1}
\end{figure}

\bigskip
\begin{table}[htp] \centering

\caption{Zn-content (\textit{x}), doping level (\textit{p}), \textit{T}$_{c}$, transition width $\Delta$\textit{T}$_{c}$ and annealing condition}
\begin{tabular}
[c]{ccccc}\hline
\textbf{Zn-content (\textit{x})} & \textbf{Doping level (\textit{p})} &
\textit{T}$_{\mathbf{c}}$ (K) & $\Delta$\textit{T}$_{\mathbf{c}}$ (K) &
\textbf{Annealing condition}\\\hline
0 & 0.16 & 93.5 & 0.5 & 500 $
{{}^\circ}
$C, 3 weeks\\
0 & 0.145 & 89 & 0.5 & 540 $
{{}^\circ}
$C, 3 weeks\\
0 & 0.137 & 81 & 2 & 580 $
{{}^\circ}
$C, 3 weeks\\
0 & 0.13 & 71 & 3 & 625 $
{{}^\circ}
$C, 2 weeks\\
0 & 0.11 & 61 & 3 & 675 $
{{}^\circ}
$C, 2 weeks\\
0.007 & 0.16 & 82 & 1 & 500 $
{{}^\circ}
$C, 3 weeks\\
0.007 & 0.137 & 64 & 4 & 580 $
{{}^\circ}
$C, 3 weeks\\
0.007 & 0.13 & 53 & 6 & 625 $
{{}^\circ}
$C, 2 weeks\\
0.007 & 0.11 & 43 & 7 & 675 $
{{}^\circ}
$C, 2 weeks\\
0.012 & 0.137 & 53 & 5 & 580 $
{{}^\circ}
$C, 3 weeks\\
0.012 & 0.13 & 37 & 5 & 625 $
{{}^\circ}
$C, 2 weeks\\
0.012 & 0.11 & 29 & 7 & 675 $
{{}^\circ}
$C, 2 weeks\\\hline
\end{tabular}
\label{TableKey}
\end{table}

\newpage
\subsection{II. Error bars}

The temperature dependent reflectivity measurements were performed with a
Bruker 80v Fourier Transform Infrared spectrometer. The optical conductivity
spectra were obtained from the measured reflectivity spectra by using the
Kramers-Kronig transformation. In Fig. S2 we plot the reflectivity and the
calculated optical conductivity spectra for the Zn-free sample at \textit{p} =
0.11 as an example up to high energy region. The following supplementary
figures are obtained from these data. In our previous work [3], we also
presented some of these data with other Zn-substituted samples, as well.

\begin{figure}[ptbh]
\centering
\includegraphics[
height=4.4421in,
width=3.5574in
]
{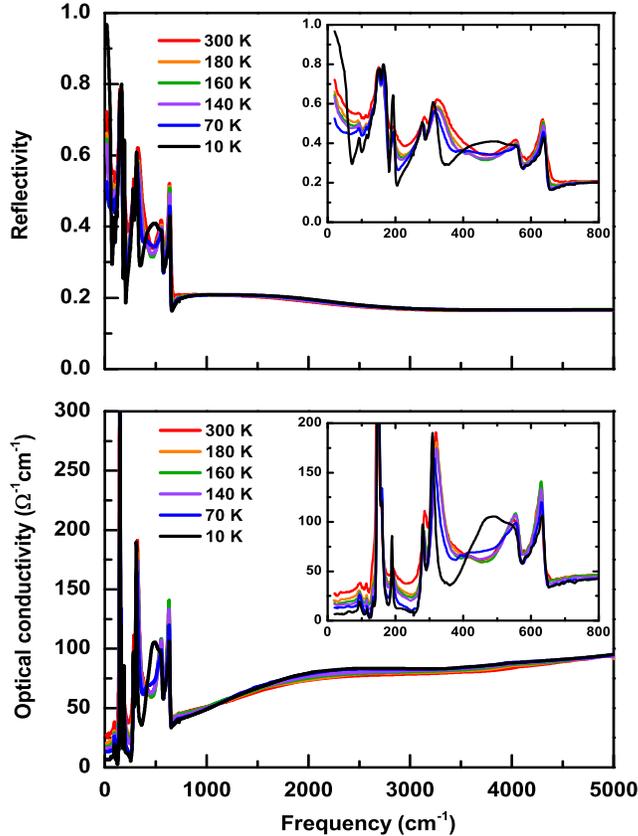}
\caption{Temperature dependent reflectivity and optical conductivity
($\sigma_{1}(\omega)$) of YBa$_{2}$Cu$_{3}$O$_{y}$ at $p=0.11$.
Insets show the lower energy part of each plot.}
\label{FS3}
\end{figure}

\bigskip

The reflectivity measurements were carried out in six different energy
regions. The spectra in the overlapped energy regions coincide with each
other. Moreover, measurement at each temperature had been repeated several
times. The repetition of the measurement at each energy region allow us to
specify an error bar in this energy range. Moreover, we can also define
another error bar in the coinciding energy regions for different measurement
ranges, which eventually gives us the overall error in our reflectivity
measurements. The error bars are calculated with the same way for each
temperature separately. As a result, the maximum error in our reflectivity
measurements is better than 0.5\%. The error bars in the superfluid densities
were estimated from the error of the conductivity that was calculated from
reflectivity. The procedure of the normal component subtraction introduces
additional errors. The error bars plotted in Fig. 2 are relative ones respect
to the data at \textit{T} just above \textit{T}$^{\ast}$.

\subsection{III. Superfluid density estimated from the real and imaginary part
of the optical conductivity}

In the superconducting state, real part of the optical conductivity
($\sigma_{1}(\omega)$) for the superconducting carriers is condensed to a
$\delta$-function at $\omega$ = 0. Then we expect to see the following
relation between the imaginary optical conductivity ($\sigma_{2}(\omega)$) and
the superfluid density ($\omega_{ps}^{2}\varpropto n_{s}/m\ast,$ $n_{s}$ is
the superfluid density) through Kramers-Kronig transformation; $\omega
\sigma_{2}(\omega)$ $\varpropto$ $\omega_{ps}^{2}$. However, in this approach
it is assumed that there is no normal carrier component in $\sigma_{1}
(\omega)$ and $\sigma_{2}(\omega)$ below the gap energy, which is not the case
for the high temperature cuprate superconductors. Therefore, in order to
estimate the correct values of $\omega_{ps}^{2}$ we have to subtract the
normal carrier component from $\sigma_{2}(\omega)$. In this supplementary, we
refer the normal carrier component as $\sigma_{2,normal}(\omega)$ that is
calculated with the Kramers-Kronig transformation of $\sigma_{1}(\omega)$. In
Figure S3, we plot our $\sigma_{2}(\omega)$ values, as well as the calculated
normal carrier components $\sigma_{2,normal}(\omega)$ for \textit{x} = 0 and
\textit{p} = 0.11 at several specific temperatures, 300 K, 180 K (just above
\textit{T}$_{p}$), 160 K (\textit{T}$_{p}$), 140 K, 70 K (just above
\textit{T}$_{c}$), and 10 K (the lowest temperature). Similar trends have been
observed for all of our samples.

We also plot the difference of $\sigma_{2}(\omega)$ and $\sigma_{2,normal}
(\omega)$ multiplied with $\omega$ that is proportional to $\omega_{ps}^{2}$,
which demonstrates the $\omega$ - constant behavior of the superfluid density.
If there is no superconducting carrier component, $\sigma_{2,normal}(\omega)$
should be identical to $\sigma_{2}(\omega)$. Indeed this behavior was observed
at high temperatures (In Fig. S3, 300 K and 180 K). When we cool down the
sample below 180 K, a difference between $\sigma_{2}(\omega)$ and
$\sigma_{2,normal}(\omega)$ appears and becomes larger with decreasing
temperature, which indicates a growth of the superconducting response in the
system. Finally, we plot our 10 K data in a wider energy range that can be
compared with the published studies [4].

\begin{figure}[ptb]
\centering
\includegraphics[
height=4.3306in,
width=3.4676in
]
{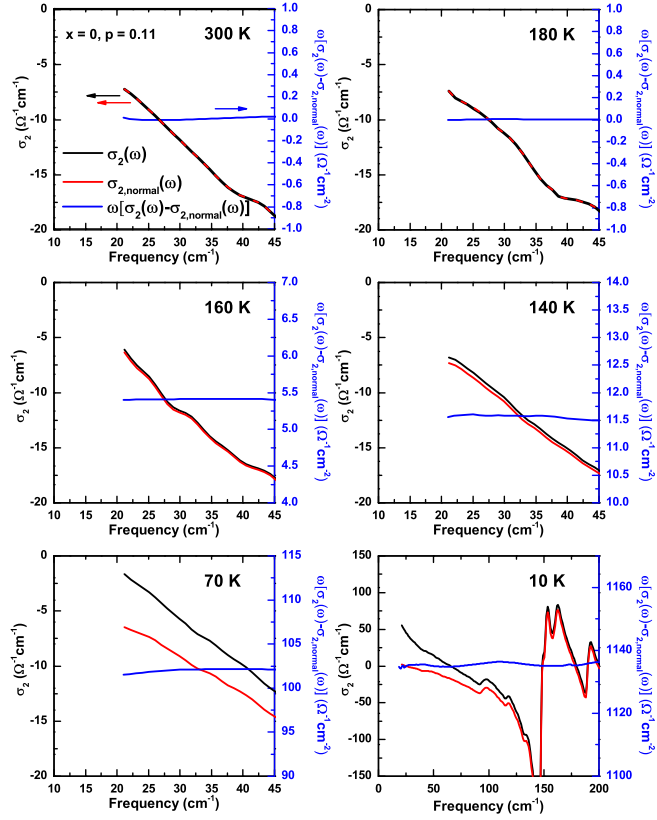}
\caption{$\sigma_{2}(\omega)$ and $\sigma_{2,normal}(\omega)$ (left axis) and
$\omega\Delta\sigma_{2}(\omega)$ (right axis) for \textit{x} = 0 at several
temperature, where $\Delta\sigma_{2}(\omega)$ = $\sigma_{2}(\omega)$ -
$\sigma_{2,normal}(\omega)$. At high temperatures down to 180 K there is no
superconducting carrier component, hence the normal carrier component is equal
to $\sigma_{2}(\omega)$. With lowering temperature we start to see difference
of the $\sigma_{2}(\omega)$ and $\sigma_{2,normal}(\omega)$ indicating the
superconducting carrier response.}
\label{FS4}
\end{figure}

\begin{figure}[ptbh]
\centering
\includegraphics[
height=2.8932in,
width=2.8932in
]
{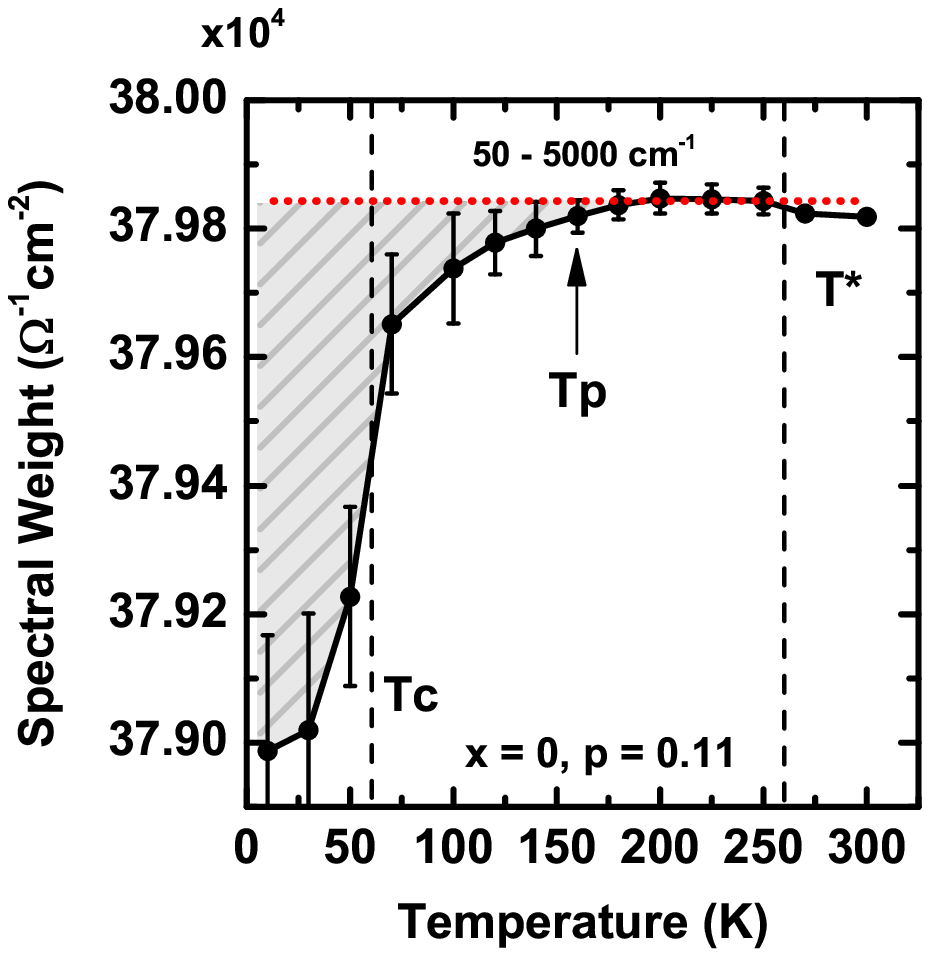}
\caption{Temperature dependent spectral weight (SW) calculated through the
equation $SW=\protect\int_{50}^{5000}\sigma_{1}(\omega)d\omega$ for \textit{x}
= 0 at \textit{p} = 0.11. \textit{T}$_{c}$ and \textit{T}$^{\ast}$ values are
marked with dashed lines. Hatched area represent the existence of the
superconducting response. We marked the \textit{T}$_{p}$ with an arrow as the
point that we start to see this response. Error bars are calculated as a
relative error to the data just above \textit{T}$^{\ast}$.}
\label{FS2}
\end{figure}

For the real part of the optical conductivity, we calculated the superfluid
density $\omega_{ps}^{2}$ from the missing area in $\sigma_{1}(\omega)$. When
the pseudogap opens the low energy spectral weight is suppressed, and the lost
spectral weight is transferred to the high energy region. In our previous work
[3], we showed that the spectral weight transfer is completed below 5000
cm$^{-1}$. Namely, the spectral weight is conserved below 5000 cm$^{-1}$. We
plot the temperature dependence of the spectral weight between 50 and 5000
cm$^{-1}$ (Fig. S4). While this spectral weight shows the temperature
independent behavior down to \textit{T}$_{p}$ (arrow in Fig. S4), below this
temperature, it starts to decrease, creating a missing area, \textit{A}, that
corresponds to the superconducting carrier response. We can calculate
$\omega_{ps}^{2}$ from this missing area by using the equation $\omega
_{ps}^{2}$ = $120/\pi A$. To compare $\omega_{ps}^{2}$ with that obtained from
$\sigma_{2}(\omega)$ we calculated the cumulative missing area with respect to
the temperature just above \textit{T}$_{p}$ at each temperature, and plotted
in Fig. 2.

\subsection{IV. Supplementary References}

[1] Y. Yamada, Y. Shiohara, Physica C \textbf{217}, 182 (1993)

[2] R. Liang, D. A. Bonn, W. N.Hardy, Physical Review B \textbf{73}, 180505(R) (2006)

[3] E. Uykur et al., J. Phys. Soc. Jpn. \textbf{82}, 033701 (2013)

[4] D. N. Basov et al., Science \textbf{283}, 49 (1999)

\end{document}